# Evaporation of Active Drops: Dynamics of Punctured Drops and Particle Deposits of Ring Galaxy Patterns

Ghansham Rajendrasingh Chandel,[1] Vishal Sankar Sivasankar,[1] and Siddhartha Das[1]*

[1]Department of Mechanical Engineering, University of Maryland, College Park, MD 20742

*Email: sidd@umd.edu

**Abstract**

Active drops refer to drops with the ability to self-migrate: these drops typically attain this ability by virtue of containing of active particles that derive energy from their environment and undergo directed motion inside the drops, thereby creating intricate stress distribution within these drops. Here we describe the evaporation dynamics of a slender active nematic drop. The stresses induced by active nematic particles present within the drop enables fascinating drop evaporation dynamics, consisting of an initial pinned stage and a late runaway stage. Unlike regular drops, during the pinned stage (for extensile drops) the drops encounter puncturing at their centers, followed by a receding motion of the newly formed inner contact line with the liquid flux pushing the (nematic) particles towards the inner and the outer contact lines: the result is the formation of a unique "ring-galaxy"-like particle deposition pattern. We identify three non-dimensional parameters, representing the activity, the aspect ratio, and the receding contact angle, which dictate the occurrence of puncturing, the overall evaporation time, and the possible deposit patterns for the extensile drops and the contractile drops. Finally, we argue that such unique evaporation and particle deposition dynamics can be leveraged for altering lifetime of drops for cooling and bio-applications and creating customized thin film deposits with potential 3D printing applications.

Active drops are liquid drops consisting of particles that derive energy from their surroundings to produce mechanical work, typically by self-propulsion or induction of active stresses [1-4]. Multiple studies have focused on understanding the fundamentals of active drop dynamics [5-18], which is a topic that have found applications in developing novel active emulsions [19-22] and droplet micro-swimmers [23-25], enhancing liquid-liquid mixing [26], explaining bio-locomotion [27], triggering molecular-scale assembly [28], providing growth models for protocells [29], enforcing spatio-temporal control of cargo delivery [30], etc.

In this letter, we shall study a hitherto unexplored topic of active drops: evaporation of an active liquid drop. Evaporation leads to a continuous mass loss. The interplay of such mass loss with the active stresses encountered by the drop (a slender drop containing active nematic particles) leads to fascinating drop dynamics and the associated particle deposition pattern formation. First, we discover the formation of punctured drops: the active stresses, in conjunction with the evaporation triggered mass loss, force a part of the air-liquid interface of the drop to descend down and hit the substrate. The result is the formation of a new contact line (or inner contact line or ICL) and a donut-shaped drop. Second, unlike regular drops, active stress constrains outer contact line (OCL) to remain pinned, while the ICL sweeps at receding contact angle due to the continuous evaporation-driven mass loss, triggering an inside-out evaporation. Note that puncturing can occur for regular non-active drops as well if the air-liquid interface hits the substrate (possibly due to uneven substrate). Third, such specific drop shape and evaporation dynamics lead to the formation of particle deposits of ring-galaxy-like pattern, characterized by an outer ring and an inner diffuse zone of particles. Such unique punctured drop dynamics and the particle deposition pattern are realized for extensile drops with certain combination of dimensionless parameters characterizing the activity, the drop aspect ratio, and the receding contact angle. Finally, a comparison with contractile and regular drops demonstrates that activity strongly controls evaporation times and can aid in manipulating contact line dynamics.

*Mathematical Framework:*

We consider evaporation of an active drop, or a drop laden with active nematic particles. Considering the evaporation to be diffusion-dominated [31-34], the equations governing the dynamics of the evaporating active drop can be expressed as (by modifying the equations for the dynamics of non-evaporating active drops [3]):

$$\partial_t h + \boldsymbol{\nabla}_\perp \cdot (\Delta h \bar{\boldsymbol{u}}_\perp) + J_z/\rho = 0, \quad \bar{\boldsymbol{u}}_\perp = \frac{(\Delta h)^2}{3\mu}(\boldsymbol{\nabla}_\perp \gamma \nabla_\perp^2 h - (\boldsymbol{\nabla} \cdot \boldsymbol{\sigma}^a)_\perp), \quad (1a,1b)$$

$$\nabla^2 c = 0, \quad \boldsymbol{J} = D\boldsymbol{\nabla} c. \quad (2a,2b)$$

Here, $h$ is the local, instantaneous drop height, $\Delta h$ is the film thickness (equal to $h$ when the substrate is at $z=0$), $\bar{\boldsymbol{u}}_\perp$ is the average in-plane velocity vector, $\boldsymbol{\nabla}_\perp$ represents the gradient in the normal plane, and $\gamma$, $\mu$, and $\rho$ are the surface tension, dynamic viscosity, and the density of the drop. Furthermore, $J_z$ is the evaporation flux in $z$ direction and $\boldsymbol{\sigma}^a = -\xi(\boldsymbol{r},t)(\boldsymbol{p} \otimes \boldsymbol{p})$ is the active nematic stress tensor [3,4,35-37] ($\boldsymbol{p}$ and $\xi$ are the polarization field and the effective concentration of the nematic particles inside the drop). $\boldsymbol{\sigma}^a$ is symmetric under 180° rotation as we lack head-tail distinction in our nematic particles (true up-to first order for polar swimmers). We consider vortex defect, i.e., $\boldsymbol{p} = \hat{\theta}$, which is one of the stable topological defects of nematic field and satisfies the anchoring boundary condition by default [3,4,38-40]. Lastly, $c$ is the water vapor concentration and $D$ is the diffusivity of water vapor in air.

We solve eqs.(1,2) under the assumption that at a given time $\xi(\boldsymbol{r},t)$ remains uniform in space [i.e., $\xi(\boldsymbol{r},t) = \xi(t)$]. We satisfy this assumption by conserving the total amount of particles inside the drop, i.e.,

$$\int_{V(t)} \xi(\vec{r},t)\,dV = \xi_0 V_0 \Rightarrow \xi(t) = \frac{\xi_0 V_0}{V(t)}, \quad (3)$$

where $\xi_0$ represents the initial concentration of the particles inside the drop (of initial volume $V_0$). Typical to evaporating drop studies, in eq. 1(b), the magnitude of $\bar{u}_\perp$ is assumed to be much smaller than the speed at which the drop relaxes when perturbed. In other words, at time scales of evaporation, the drop appears quasi-static (i.e., in stress equilibrium at all times). Under such conditions, and assuming axisymmetry, we can reduce eq. 1(b) to obtain the time independent stress balance as [41]:

$$\gamma \frac{\partial}{\partial r}\left[\frac{1}{r}\frac{\partial}{\partial r}\left(r\frac{\partial h}{\partial r}\right)\right] + \frac{\xi(t)}{r} = 0. \quad (4)$$

Eq.(4) can be solved analytically in presence of the condition $h(r = R(t), t) = 0$ [$R(t)$ is the drop radius, see Fig. 1(i)] yielding [41]:

$$h(r,t) = -\frac{\xi(t)}{4\gamma}\left[r^2 \ln\left(\frac{r}{R(t)}\right) + (R(t)^2 - r^2)\right] + \frac{\Theta(t)}{2R(t)}(R(t)^2 - r^2) + C(t)\ln\left(\frac{r}{R(t)}\right). \quad (5)$$

To satisfy eq. 1(a), constants of integration in eq. (5) [$\Theta(t), C(t)$] as well as $R(t)$ must be functions of the evaporation time scale. Using lubrication approximation and assuming initial pinned-contact-line evaporation stage, we can employ the volume conservation condition as [this condition will help us to evaluate $\Theta(t)$] [31]:

$$\frac{dV(t)}{dt} = \frac{d}{dt}\int_{r_{in}}^{R} h(r,t) 2\pi r\, dr = \dot{V}. \quad (6)$$

For the pinned CL evaporation stage, $r_{in} = 0$ and $R(t) = R$. The condition that $h(0,t)$ is finite during the pinned stage enforces $C(t) = 0$. Under such circumstances, and using eq.(3) that provides $\xi(t)$, eq.(5) can be re-written to provide the dimensionless drop height, $H$, as [41]:

$$H(a,T) = -\frac{\Lambda}{1-T}\left(a^2 \ln(a) + (1-a^2)\right) + \left(\frac{\Omega}{2}(1-T) + \frac{3\Lambda}{4}\frac{1}{1-T}\right)(1-a^2). \quad (7)$$

Here $H = h/R$, $T = -t\dot{V}/V_0$, $a = r/R$, $\Omega = 4V_0/\pi R^3$ (aspect ratio of the drop), and $\Lambda = \xi_0 R/4\gamma$ (dimensionless activity). This equation is valid until the time $T < T_P$, however, eq.(7)

shows that for extensile drops (i.e., drops with $\Lambda > 0$), for $T > T_P$ ($T_P = 1 - \sqrt{\Lambda/2\Omega}$ is the dimensionless puncturing time, which indicates the onset of drop puncturing), $H(a = 0, T) < 0$ (i.e., the air-liquid interface of the drop near center goes below the substrate) and hence the solution no longer holds. We must resolve this issue as the solution must remain physical until the drop evaporates to zero volume.

According to contact-angle hysteresis (CAH) model proposed by [42,43], contact-lines with contact-angles $\theta$, such that $\theta_R < \theta < \theta_A$ ($\theta_R$ and $\theta_A$ are receding and advancing contact angles), remain pinned. Eq.(7) shows that at $T = T_P$, the drop surface (air-liquid interface) touches the ground (at $a=0$, i.e., at the drop center) tangentially (*this is the drop puncturing event*), and hence the contact angle (made by the ICL caused by puncturing) is zero. This is energetically unfavorable under CAH model and the contact-line at the center must follow receding motion until $\theta(a_P, T_P^+) = \theta_R$ ($a_P$ is the non-dimensional puncturing radius of the drop) [Fig. 1(c)]. Since we assumed slow, diffusion-driven evaporation, the drop will attain its new equilibrium position almost instantaneously after the puncturing, and we represent this time (non-dimensionally) as $T_P^+$. This gives us our complete set of boundary conditions, $\theta(a_{in}(T), T_P^+) = \theta_R$ [$a_{in}$ is the dimensionless radius of the receding inner contact line, see Fig. 1(e)], $H(a_{in}(T), T) = 0$, and $H(1, T) = 0$: these conditions can be used to solve eq. 1(b) (a 3$^{rd}$ order ODE) to obtain the drop dynamics *post puncturing* [eq.(7) gives the analytical solution for the drop profile *before puncturing*]. We cannot impose any condition on contact angle at the OCL, which is always pinned [with $\theta(1, T) > \theta_R$] for a quasi-static extensile drop.

We can only have a semi-analytical solution for the drop profile after puncturing ($T > T_P$) as we cannot find an analytical solution to eq.(2) for a punctured geometry. Furthermore, the equations we get by substituting our boundary conditions (see above) in eq.(5) [eq.(5) with $C(t) \neq 0$ can represent the dynamics of a punctured drop] are not algebraic and their representation using standard functions is not possible. Therefore, we find $\Theta(t)$ and $C(t)$ numerically (see [41] for details) such that the boundary conditions are satisfied along with eqs.(2a,2b,6). To get a unique solution, we must consider the conservation of particles, i.e., account for the rate of deposition of active particles at inner and outer contact lines to determine $\xi(t)$: this, of course, will necessitate solving the species transport equation for the polarization field $\boldsymbol{p}$ [44], which may depend on the exact geometry of active particles. Without doing such calculations, similar to [45] for receding drops, we make a simplifying assumption that the density of the non-deposited particles (or particles inside the drop) does not change after puncturing, i.e., $\xi(T \geq T_p) = \xi(T_p)$, i.e., we assume that the reduction in the drop volume is followed by an appropriate deposition of the (active) particles. This gives us a neat method to probe the post-puncturing drop physics without introducing new complexities (see [41] for the numerical method).

*Dynamics of puncturing-vs-non-puncturing extensile drops*

Figs. 1(a,b) show the time-dependent evaporation dynamics of an extensile drop corresponding to a condition (i.e., a specific combination of $\Lambda$, $\Omega$, and $\theta_R$) that ensures drop puncturing. Fig. 1(a) shows the drop profile for the conditions when the contact line (CL) remains pinned during the evaporation. The very last profile in Fig. 1(a) denotes the profile (at $T = T_p$) that corresponds to the onset of puncturing that forms an inner contact line (ICL) [Fig. 1(e)]. Post puncturing, under CAH, the active stress ensures that $\theta$ at the ICL becomes equal to a critical value ($\theta_R$) at $T = T_P^+$ and the corresponding drop profile (from which the ICL starts to recede at evaporation time scales) is the first profile in Fig. 1(b). Fig. 1(c) shows this instantaneous transition. Post-puncturing, therefore, the motion of the drop [of "doughnut" shape, see Fig. 1(e)], occurs with the ICL receding

towards the pinned outer contact line (OCL). Fig. 1(b) shows the time evolution of these drop profiles post puncturing. Figs. 1(d,e) provide the 3D profiles corresponding to the separate drop profiles at pre-puncturing (pinned CL) and post-puncturing (receding ICL, but pinned OCL) stages. Figs. 1(f,g) provide the evaporating drop profiles for $\Lambda$, $\Omega$, and $\theta_R$ values that ensure that there is no puncturing for the extensile drop. Fig. 1(f) shows the profile for the drop undergoing evaporation with pinned CL. This leads to a progressive decrease in the contact angle eventually leading to the contact angle to be equal to $\theta_R$ enforcing a subsequent evaporation with the CL receding towards the drop center [the corresponding profiles are shown in Fig. 1(g)]. In the absence of puncturing, the ICL does not form; hence the receding CL motion ensures a progressive and axisymmetric shrinkage of the drop [see 3D images in Figs. 1(h,i)]. Finally in Fig. 1(j), we provide the parametric phase space that determines whether or not the active and extensile evaporating drop will undergo puncturing: larger $\Lambda$ and $\Omega$ and smaller $\theta_R$ seem to promote drop puncturing.

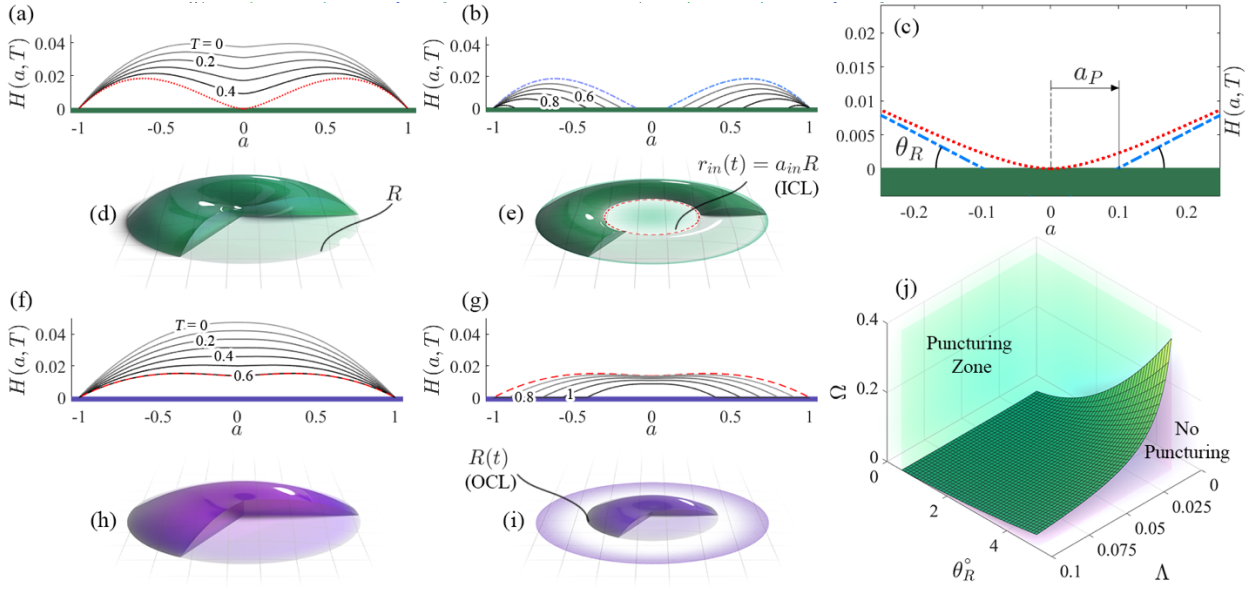

**Figure 1.** (a,b) Time evolution of the profiles of an evaporating extensile nematic drop undergoing puncturing for (a) time instants (pre-puncturing) when the CL is pinned and (b) time instants (post-puncturing) when the puncturing-induced ICL recedes but the OCL remains pinned. (c) Transition from the profile at the onset of puncturing [last profile in 1(a); $a_p$ is the dimensionless puncturing radius] to the profile where $\theta = \theta_R = 3°$ and the ICL starts to recede [first profile in 1(b)]. (d,e) 3D drop profiles corresponding to (d) a pre-punctured drop profile (T=0.5) and (e) a post-punctured drop profile (T=0.65). In (e), dimensionless and time dependent inner radius, $a_{in} = r_{in}/R_0$ is noted. In (a-e), we consider $\Lambda$=0.05, $\Omega$ =0.1, $\theta_R$=3°. (f,g) Time evolution of the profiles of an evaporating extensile nematic drop undergoing no puncturing for (f) time instants when the contact line is pinned and (g) time instants when the contact line recedes. In (f,g), the drop profile demarcating the transition between pinned and receding contact line cases is shown by a dashed line. In (a,b,f,g), the dimensionless times associated with the drop profiles have been identified. (h,i) 3D drop profile corresponding to the no-puncturing case for (h) T=0.603 (pinned CL case) and for (i) T=1 (receding CL case). In (f-i), we consider $\Lambda$=0.01, $\Omega$ =0.1, $\theta_R$=3°. (j) Phase space that identify whether or not the evaporating extensile nematic drop will undergo puncturing or not.

*Formation of ring-galaxy-like deposits*

Puncturing of the drop leads to the formation of two separate CLs. As a result, there will be evaporation-driven liquid fluxes in two separate directions (towards the two separate CLs): Fig. 2(a) provides the direction and magnitude of these fluxes at a time after puncturing for the case studied in Figs. 1(a-e). The fluxes at the ICL and the OCL approach $-\infty$ and $+\infty$ respectively for $T > T_p$, suggesting particles just adjacent to the CLs are strongly pulled towards the CL by the evaporation-driven mass loss (see [41] for flux distributions for other parameters). Fig. 2(b) shows the evaporative flux, $J_z$, which induces the liquid fluxes: the maximum evaporative fluxes occur at the locations of the ICL and the OCL with the ICL progressively receding towards the OCL, which corresponds to increasing $a_{in}(T)$ [Fig. 1(e) defines $a_{in}(T)$]. Under these circumstances, the active particles deposit as a single "ring" at the OCL, but they spread out as a diffuse and continuous band with progressively decreasing concentration in the direction of motion of the ICL. The eventual result, therefore, is a ring galaxy like deposit [Fig. 2(c)]. The relative concentration of the deposited particles constituting the ring and the diffuse zone depends on (i) the time of the drop puncturing (since quicker the puncturing, greater will be the timespan for which the ICL exists and recedes, thereby leading to a larger concentration of the particles in the diffuse zone), (ii) the radius of the ICL immediately after puncturing, and (iii) the speed at which the inner contact line recedes (lesser this speed greater will be the time that the particles have to get deposited). Fig. 2(d) provides the numerical results for the trajectory of the ICL [variation of $a_{in}$-vs-$T$] as functions of $\Lambda$ and $\Omega$. From this figure, one can also obtain the ICL receding velocity (by estimating the slope). The variation of $a_p$ and the corresponding drop puncturing time ($T_P$) (see inset), as functions of $\Lambda$ and $\Omega$, are separately plotted in Fig. 2(e). Finally, Fig. 2(f) provides the schematic of the expected ring-galaxy deposition pattern for different combinations of $\Lambda$ and $\Omega$.

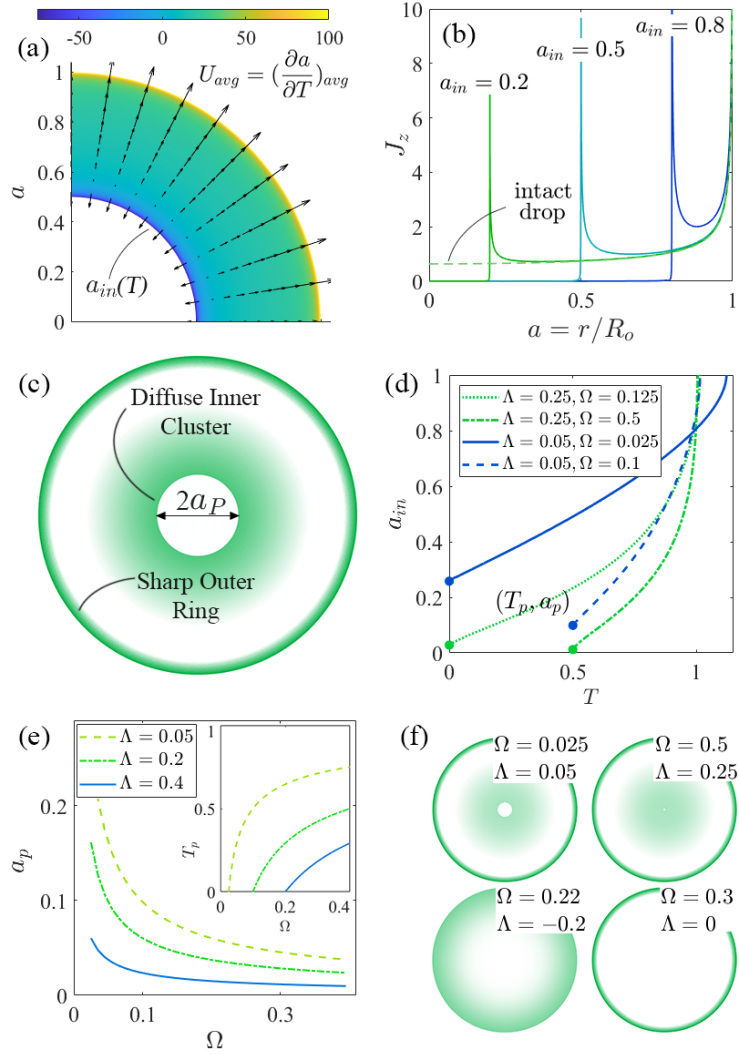

**Figure 2.** (a) Top view of the punctured drop showing the liquid fluxes [quantified in terms of the dimensionless velocity field $U_{avg} = \left(\frac{\partial a}{\partial T}\right)_{avg}$] directed towards the ICL and the OCL. Results are shown for $\Lambda$=0.25, $\Omega$ =0.50, $\theta_R$=3°, and $T$=0.95. (b) Flux of vaporized liquid corresponding to different locations of the receding ICL (which correspond to different values of $a_{in}$) and fixed OCL. (c) Schematic of the ring-galaxy like particle deposit pattern. (d) Variation of $a_{in}$-vs-$T$ for different $\Lambda$ and $\Omega$. Non-dimensional puncturing time ($T_p$) and puncturing radius ($a_p$) [$a_p = (a_{in})_{T=T_p}$] has been indicated for one of the curves. (e) Variation of $a_p$ and $T_p$ (see inset) with $\Lambda$ and $\Omega$. (f) Anticipated ring-galaxy deposition patterns for different combinations of $\Lambda$ and $\Omega$.

*Extensile-versus-Contractile Drops*

In Figs. 1 and 2, we have considered the dynamics of an extensile drop and the expected particle deposition pattern resulting from the evaporation of such drops. Fig. 3(a) provides the time evolution of the drop profiles for the contractile drop (characterized by negative activity, i.e., $\Lambda<0$). There is no puncturing of the drops and the evaporation-driven drop dynamics first occurs with the CL pinned, followed by receding CL. Contractile drops undergo much slower evaporation (evaporation time can be up to 50% more) as compared to extensile drops. Fig. 3(b) shows the dimensionless evaporation time, $T_{evap}$ (time needed for the drop volume to reach zero) as functions of $\Lambda$ and $\Omega$. For extensile drops that puncture, the evaporation time is significantly small, i.e., much smaller than the non-active drops ($\Lambda=0$) with the same aspect ratio. On the other hand, for contractile drops ($\Lambda<0$), $T_{evap}$ is significantly higher with $T_{evap}$ increasing with $-\Lambda$. The contractile drop contracts and pulls the liquid away from the three-phase contact line and towards the drop center. This reduces the contact angle forcing the contact angle to reach $\theta_R$ much earlier, thereby enforcing a quicker onset of the drop dynamics with receding CL (i.e., the drop wetted radius, $R$, starts to reduce much earlier for the contractile drops). Corollary to this effect, the depinning is delayed by positive, or extensile activity. Smaller $R$ will lead to a smaller rate of evaporation [see eq.(6)], enforcing a greater evaporation time for contractile drops as compared to the extensile and non-active drops.

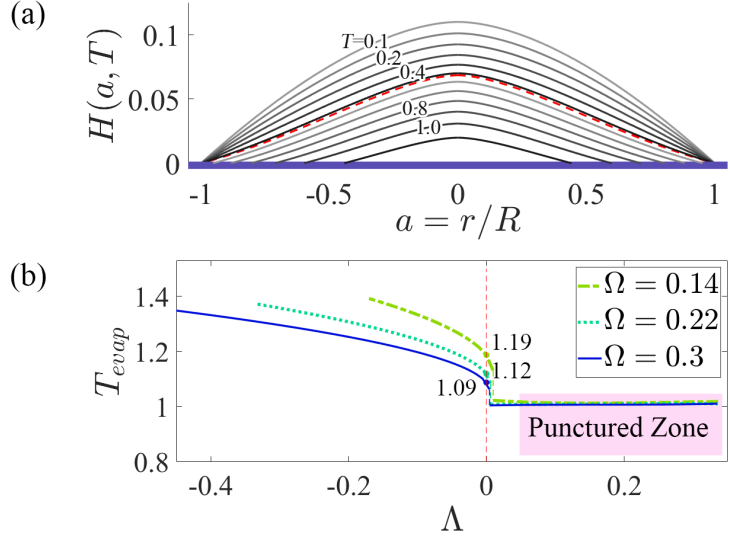

**Figure 3.** (a) Time evolution of the profiles of an evaporating contractile drop ($\Lambda = -0.04$, $\Omega = 0.2$, $\theta_R = 3°$). At $T=0.53$, the contact angle becomes equal to $\theta_R$ (profile shown by dashed line). For $T<0.53$, the drop dynamics occur with pinned CL, while for $T>0.53$, the drop dynamics occurs with receding CL. (b) Variation of the dimensionless evaporation time, $T_{evap}$, for different values of $\Lambda$ and $\Omega$. Cases with $\Lambda < 0$ correspond to contractile drops. $T_{evap}$ for extensile drops that undergo puncturing and non-active drops ($\Lambda = 0$) have been separately indicated.

In summary, we have developed a framework to probe the evaporation dynamics of extensile and contractile nematic (active) drops. The dynamics is characterized by the formation of punctured drops, the resulting ring-galaxy-like deposits of the nematic particles present within the active drop, and a wide variation in the drop evaporation lifetime. These findings can be significant for various applications. For example, the unique ability to control the drop life time by changing the nature of the activity (extensile *versus* contractile) can be leveraged for applications ranging from slow evaporation driven crystal growth [46] to fast evaporation driven rapid cooling [47,48] and faster stimulation of bacterial osmoregulation for cell viability assessment [49]. On the other hand, the ability to ensure ring-galaxy-like deposition pattern will enable self-assembly driven fabrication of components (e.g., superhydrophobic surfaces [50], 3D nanostructures [51], photonic crystals [52], etc.) that can have a controllable spatial gradient in the distribution of the constituent materials. Finally, the desired extent of particle deposition (enabled by the ability to trigger ring-galaxy like patterns), when coupled with the controllable migration ability of the active drops [4], can potentially be used for a continuous deposition of desired amount materials at desired locations paving the way for designing novel deposition and 3D printing techniques.